\documentclass[letterpaper,twocolumn,english,aps,prb,groupedaddress,superscriptaddress,showpacs,amsfonts]{revtex4-1}

\usepackage{epsfig}
\usepackage{epstopdf}
\usepackage{graphicx}
\usepackage{dcolumn}
\usepackage{bm}

\DeclareMathAlphabet{\mathpzc}{OT1}{pzc}{m}{it}


\newcommand{\beq}{\begin{equation}}
\newcommand{\eeq}{\end{equation}}
\newcommand{\bea}{\begin{eqnarray}}
\newcommand{\eea}{\end{eqnarray}}

\begin{document}

\title{Dominant superconducting fluctuations in the one-dimensional extended Holstein-extended Hubbard model}
\author{Ka-Ming Tam}
\altaffiliation{Present address: Department of Physics and Astronomy, Louisiana State University, Baton Rouge, LA 70803}
\affiliation {Department of Physics, Boston University, 590 Commonwealth Ave., Boston, MA 02215}

\author{Shan-Wen Tsai}

\affiliation{Department of Physics and Astronomy, University of California, Riverside, CA 92521}
\author{David~K.~Campbell}

\affiliation {Department of Physics, Boston University, 590 Commonwealth Ave., Boston, MA 02215}

\date{\today}

\begin{abstract}

The search for realistic one-dimensional (1D) models that exhibit dominant superconducting (SC) fluctuations effects has a long history. \cite{Bourbonnais_review,Lang_review,Ishiguro1,Ishiguro2} In these 1D systems, the effects of commensurate band fillings--strongest at half-filling--and electronic repulsions typically lead to a finite charge gap and the favoring of insulating density wave ordering over superconductivity. Accordingly, recent proposals \cite{Clay1, Clay2} suggesting a gapless metallic state in the Holstein-Hubbard (HH) model, possibly superconducting, have generated considerable interest and controversy, with the most recent work demonstrating that the putative dominant superconducting state likely does not exist. \cite{Clay2, Tam_hh,Tam_LL}
In this paper we study a model with {\it non-local} electron-phonon interactions, in addition to electron-electron interactions, this model unambiguously possesses dominant superconducting fluctuations at half filling in a large region of parameter space. Using both the numerical multi-scale functional renormalization group for the full model and an analytic conventional renormalization group for a bosonized version of the model, we demonstrate the existence of dominant superconducting (SC) fluctuations. 
These dominant SC fluctuations arise because the spin-charge coupling at high energy is weakened by the non-local electron-phonon interaction and the charge gap is destroyed by the resultant suppression of the Umklapp process. The existence of the dominant SC pairing instability in this half-filled 1D system suggests that non-local boson-mediated interactions may be important in the superconductivity observed in the organic superconductors.  

\end{abstract}
\maketitle

\section{Introduction and Background}

Many interesting novel electronic materials, including charge-transfer solids and conducting polymers, are quasi-one-dimensional physical systems and exhibit clear effects of both electron-phonon (e-ph) and electron-electron (e-e) interactions. Accordingly, a large number of one dimensional (1D) Hamiltonians have been studied as microscopic models for these systems, with considerable success in describing the range of ground states, nonlinear excitations, and optical and transport properties of these materials.
One outstanding challenge for theorists has remained, however: namely, the identification of a 1D model, incorporating both electron-phonon and electron-electron interactions with {\it realistic} interaction parameters that exhibits dominant superconducting fluctuations. This has proven especially difficult because of the strong tendency in these 1D systems to form insulating density waves ground states: charge density wave (CDW),  spin density wave (SDW), or bond-order wave (BOW). Hence the recent suggestion  \cite{Clay1}  that the 1D Holstein-Hubbard (HH) model at half-filling might have, for a narrow window of parameters, a region of dominant superconducting fluctuations has generated considerable interest. Although some later analyses \cite{Clay2, Tam_hh, Tam_LL} suggest that this model likely does not have {\it dominant} superconducting fluctuations, the initial results  \cite{Clay2} provided support for the existence of a novel gapless metallic phase intermediate between the familiar "Peierls" (CDW insulating) and the "Mott" (SDW insulating) phases. This result, tantalizingly close to the long-sought dominant SC state, has rekindled interest in finding a 1D model which does have clearly dominant SC fluctuations for a range of physically reasonable parameters. 

In this paper, we define and study such a model, which we call the "extended Holstein-extended Hubbard" (EHEH) model. We will define the model precisely in the ensuing section, but for now we note simply that it involves both on-site ($U$) and nearest-neighbor ($V$) repulsive e-e interactions and {\it non-local} e-ph interactions.

Before describing the EHEH model in detail, let us first briefly summarize the results in the Holstein-extended Hubbard (HEH) model, as these serve as a very useful guide to the surprising subtleties encountered in studies of this seemingly simple model. The Hamiltonian for the HEH model is

\begin{eqnarray}
H_{HEH} &=& - t \sum_{i,\sigma}(c_{i+1,\sigma}^{\dagger}c_{i,\sigma} + H.c.) + \omega_{0} \sum_{i} a_{i}^{\dagger}a_{i} \nonumber\\
  &+& U\sum_{i}n_{i,\uparrow}n_{i,\downarrow} + V\sum_{i}n_{i}n_{i+1} \nonumber\\
  &+& g_{\rm ep}\sum_{i,\sigma}(a_{i}^{\dagger}+a_{i})n_{i,\sigma}.
 \label{HEH}
\end{eqnarray}
Here electrons move in a tight-binding 1D lattice, where $c_{i,\sigma}^{\dagger}$ ($c_{i,\sigma}$) creates (annihilates) an electron at site $i$ with spin $\sigma$, $n_{i\sigma}$ is the electron number operator, $n_{i} = n_{i,\uparrow} + n_{i,\downarrow}$, and $a_i^{\dagger}$ creates a (dispersionless Einstein) phonon of frequency ${\omega_0}$ at site $i$. The electron density is coupled locally with strength $g_{ep}$ to phonons on the same site. The e-e interactions are described by the standard extended Hubbard model with on-site interaction $U$ and nearest neighbor interaction $V$. 

To recall the results for this model we begin with the case $g_{\rm ep}=0$, so the model reduces to the familiar extended Hubbard model, \cite{ext-Hubbard} which incorporates only e-e interactions. We focus on the half-filled case in the physical region with $U$ and $V$ both positive. For $2V>>U$, CDW fluctuations are dominant and the ground state is indeed a long-range ordered CDW. In contrast, for $2V<< U$, spin density wave fluctuations are dominant and the ground state is an (algebraically decaying) SDW. Recent results \cite{Nakamura1, Nakamura2, Sengupta, Sandvik, TF1, TF2, Zhang, Tam_EH}  have shown that there also exists a small region around the line $U=2V$ in the $U-V$ plane in which the ground state is a long-range ordered BOW  (also sometimes called a bond-charge density wave). But there is no evidence for dominant superconducting fluctuations in this model for half-filling with $U$ and $V$ positive. This result coincides with the naive expectation that for purely repulsive interactions, superconducting pairing should not exist. For the pure Hubbard model ($V=0$)the exact solution \cite{Lieb} shows that the charge gap opens immediately for $U>0$ but that it is exponentially small for small $U$.

We consider next the case of e-ph interactions only, in which we recover the equally familiar 1D Holstein model. \cite{Holstein} In a pioneering study nearly three decades ago, Hirsch and Fradkin \cite{Hirsch_Holstein} argued that for spin 1/2 electrons, in the half-filled Holstein model a charge gap opens unconditionally for any non-zero e-ph coupling and at any finite phonon frequency. In the adiabatic limit ($\omega_0=0$) the gap is also exponentially small in $g_{\rm ep}$. While some subsequent studies have supported this results, \cite{Caron, Bourbonnais, Schmeltzer}  others  \cite{Wu, Jeckelmann, Takada1, Takada2, Hotta, Zhao} have suggested that for sufficiently large ${\omega_0}$ a finite value of $g_{\rm ep}$ is required to open the gap, so for sufficiently small $g_{\rm ep}$ there is a gapless phase. 

Similarly, in the full HEH model, although the initial study \cite{Hirsch_Ext_Hubbard} suggested no gapless phase,  more recent studies \cite{Clay1,Clay2,Takada1,Takada2,Tezuka1,Tezuka2,Feshke1,Feshke2,Feshke3,Ejima,Chatterjee,Hohenadler}, have found the "gapless" metallic phase mentioned above in a small region of the $g_{\rm ep}$-$U$ plane around the line $U= 2g_{\rm ep}^2 /{\omega_0}$.  
Ref[\onlinecite{Clay2}] gives a plausible intuitive interpretation of this phase, as well as a clear statement of the important caveat of the likely exponentially small nature of any gap in this region makes drawing definitive conclusions from these numerical studies very difficult. A further reflection of the subtleties in this problem is the fact that other studies, including some recent results, do not find this gapless phase. \cite{Yonemitsu, Bindloss, Bakrim}

Whatever the ultimate resolution of this issue in the HH model, \cite{Payeur} the "fragility" of the gapless metallic state and the absence of dominant superconducting fluctuations in that model strongly suggest that we seek a more robust model in which it is clear that there are dominant superconducting fluctuations. A hint as to what sort of model could produce this result comes from considering a problem simpler than the full many-body problem: namely, the two-electron "bipolaron" problems. Recently, several groups have studied this relatively more manageable problem in models with {\it long-range} e-ph interactions. \cite{Bonca,Alexandrov,Spencer} One of the main differences between the Holstein model, which has only on-site e-ph interactions, and models with non-local or long-range e-ph interaction is that in the latter models the mass of the bipolaron is reduced considerably \cite{Bonca,Alexandrov,Spencer} with the possible enhancement of superconductivity. If these results translate to the many-electron problem, then one should expect {\it non-local} e-ph interactions to enhance the possibility of superconductivity and lead to dominant superconducting fluctuations in a large region of parameter space. Accordingly, in the ensuing sections, we study a model--the "extended Holstein-extended Hubbard" (EHEH) model-- which includes both non-local e-ph interactions and the e-e interactions of the extended Hubbard model. We shall see that this model does indeed allow for dominant SC fluctuations in a substantial portion of the $U-V$ plane. 

\section{The EHEH Model and the MFRG method}

The explicit form of the extended Holstein-extended Hubbard (EHEH) model is given by the Hamiltonian
\begin{eqnarray}
H_{EHEH} &=& - t \sum_{i,\sigma}(c_{i+1,\sigma}^{\dagger}c_{i,\sigma} + H.c.) + \omega_{0} \sum_{i} a_{i}^{\dagger}a_{i} \nonumber\\
  &+& U\sum_{i}n_{i,\uparrow}n_{i,\downarrow} + V\sum_{i}n_{i}n_{i+1} \nonumber\\
  &+& g_{\rm ep}\sum_{i,\sigma}[(a_{i}^{\dagger}+a_{i})+(a_{i+1}^{\dagger}+a_{i+1})]n_{i,\sigma}, 
\end{eqnarray}
The notation is the same as in eqn. \ref{HEH}.
In the remainder of the paper, we will measure all energies in units of $t$. This model is similar the "extended Holstein-Hubbard" model studied in the case of bipolarons. \cite{Bonca}

Working in the path-integral representation and tracing out the phonon fields, we can express the action of the EHEH model as,
\begin{eqnarray}
S &=& \int_{\sigma,k}\psi_{k}^{\dagger}(i\omega-\vec{k})\psi_{\sigma,k}\nonumber\\
  &+& \int_{\sigma,\sigma{'},\{k\}} g(k_{1},k_{2},k_{3},k_{4}) \psi^{\dagger}_{\sigma^{'},k_{4}} \psi^{\dagger}_{\sigma,k_{3}} \psi_{\sigma,k_{1}} \psi_{\sigma^{'},k_{2}},
\end{eqnarray}
where the coupling function $g$ is given by 
\begin{eqnarray}
g(k_{1},k_{2},k_{3},k_{4}) = U + 2Vcos(\vec{k}_{3}-\vec{k}_{1}) \nonumber\\
-\frac{2g_{\rm ep}^2\omega_{0}}{[\omega_{0}^2 +(\omega_1 - \omega_3)^2]} [1+cos(\vec{k}_{3}-\vec{k}_{1})]
\label{g}
\end{eqnarray}
To analyze this Hamiltonian, we will use our version of the  multi-scale functional renormalization group (MFRG) method. \cite{Tsai1,Tsai2} This method has been used successfully to study a number of problems \cite{Tam_hh,Tam_LL, Tam_EH} in models involving both e-ph and e-e interactions because it captures not only the electron correlation effects but also the coupling of different energy scales induced by phonon retardation effects. In its simplest form, the MFRG can be used to follow the flows of the various couplings as the energy cut-off is lowered. However, it has been shown that retardation can modify the various scalings between different correlation functions for the Luttinger liquid. \cite{Loss,Chen,Tam_LL} Therefore, instead of studying the couplings, we use the MFRG to study directly the various {\it susceptibilities} which take the contributions from different energy scales into account. \cite{Tam_EH} The dominant ordering can be inferred from the flows of the susceptibilities corresponding to different order parameters as the cutoff is lowered. 

In the ensuing section, we calculate the susceptibilities corresponding to the charge density wave (CDW), the spin density wave (SDW), and both the the singlet (SS) and triplet (TS) superconducting order parameters. The dominant instability among them is determined by the most divergent susceptibility as the cut-off, $\Lambda$, is lowered. 
For completeness, we present in the Appendix a short summary of the MFRG method and the scaling equations for the susceptibilities. For further details, readers are referred to several previous articles. \cite{Tsai1,Tsai2,Tam_hh,Tam_EH} In the next section, we turn directly to the results of our MFRG study.

\section{Results for the Susceptibilities Flows}

\begin{figure}[]
\centerline{
\includegraphics*[height=0.26\textheight,width=0.36\textwidth, viewport=0 0 360 300,clip]{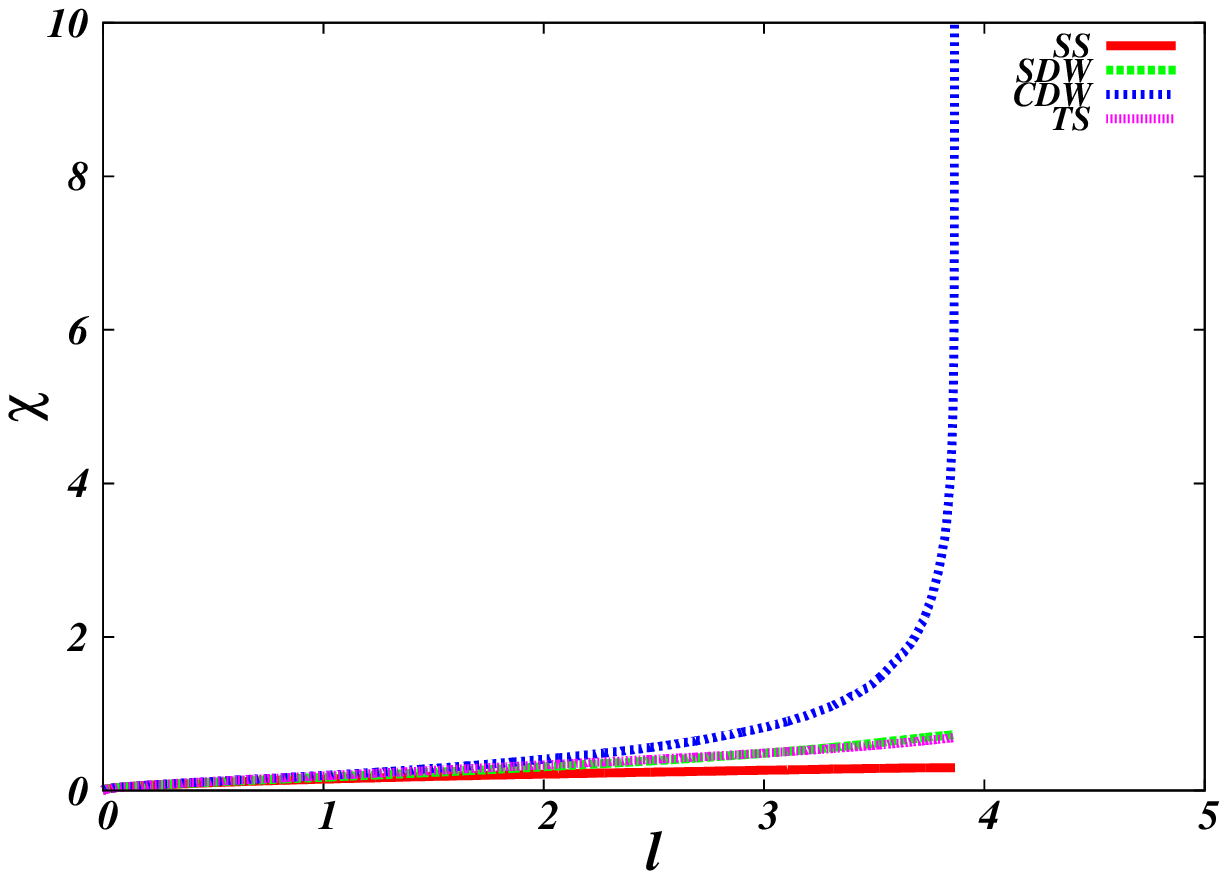}} 
\centerline{
\includegraphics*[height=0.26\textheight,width=0.36\textwidth, viewport=0 0 360 300,clip]{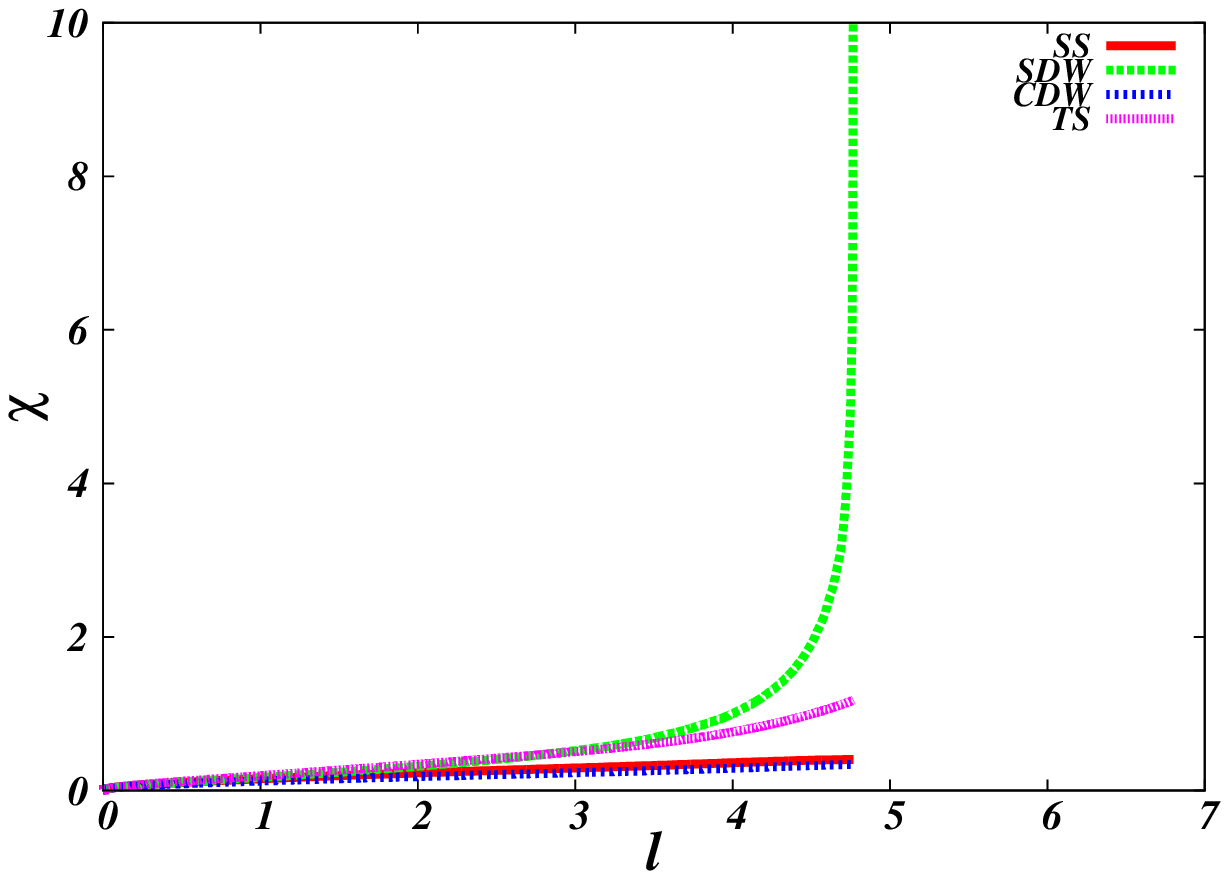}}
\centerline{
\includegraphics*[height=0.26\textheight,width=0.36\textwidth, viewport=0 0 360 300,clip]{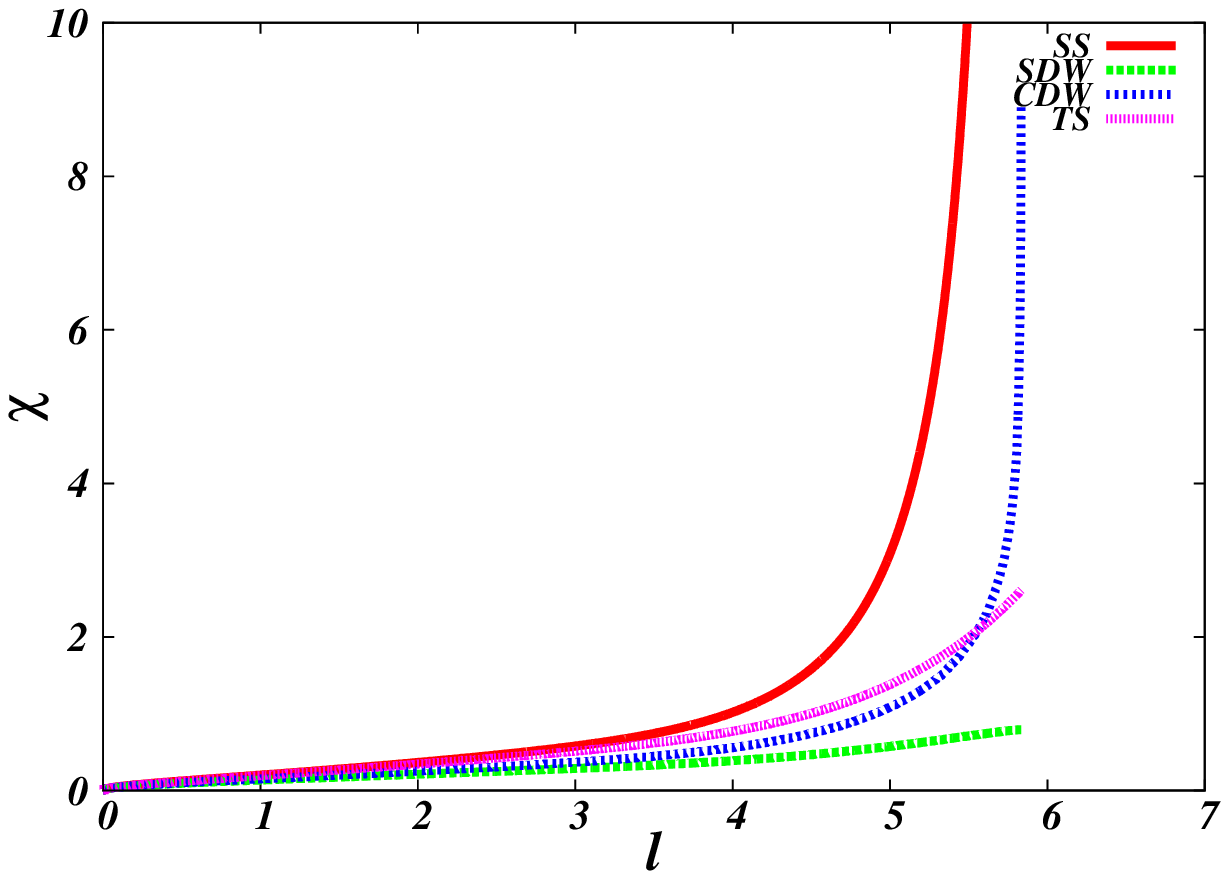}}
\caption{Susceptibility flows for $\omega_{0}=1.0$, $g_{ep}=0.5$ with various electron-electron interactions $U$ and $V$. $U=1.0, V=0.6$ for the top panel, $U=1.0, V=0.2$ for the middle panel, and 
$U=0.2, V=0.2$ for the bottom panel.
 }
\label{susceptibilities_flows}
\end{figure}

In Fig. \ref{susceptibilities_flows} we show the susceptibility flows for three different sets of e-e interaction parameters, for fixed e-ph coupling and phonon frequency. In the top panel of Fig. \ref{susceptibilities_flows}, we choose values of $U$ (=1.00) and $V$ (=0.60) which are in the region where, for the extended Hubbard model {\it without} phonons, the dominant
susceptibility is known to be a CDW. \cite{Tam_EH} Our calculations show that this result is unchanged by the inclusion of phonons for the values of $g_{ep}$ and $\omega_0$ shown in the figure. 

In the middle panel, we show the flows for $U=1.00$ and $V=0.20$ in the region where, for the  extended Hubbard model {\it without} phonons, the dominant susceptibility is known to be a SDW. \cite{Tam_EH} Again, we find that this result is unchanged by the inclusion of phonons for the values of $g_{ep}$ and $\omega_0$ shown in the figure. 

Finally, in the bottom panel of Fig.  \ref{susceptibilities_flows}, we tune down the e-e interaction to $U=0.20$ and $V=0.20$. In the extended Hubbard model {\it without} phonons, this would be in the CDW region. Instead, we find that for the values of the phonon parameters in Fig. \ref{susceptibilities_flows}, the superconducting susceptibility clearly becomes dominant. This provides an in-principle proof that non-local e-ph interactions are able to drive the pairing instability even at half-filling.  

\begin{figure}[]
\centerline{
\includegraphics*[height=0.28\textheight,width=0.39\textwidth, viewport=0 0 360 280,clip]{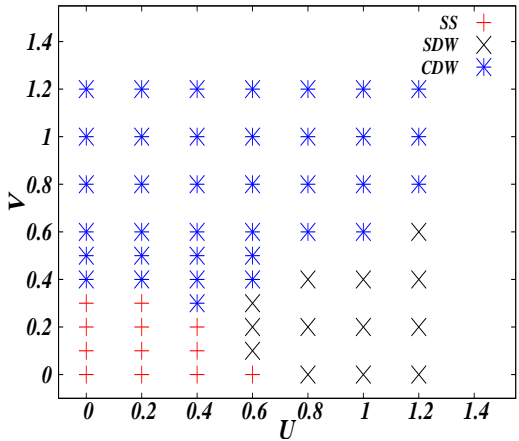}} 
\caption{Phase diagram for $g_{ep}=0.5$ with $\omega_{0}=1.0$.}
\label{phase_diagram}
\end{figure}

In Fig. \ref{phase_diagram} we display the "phase diagram" in the $U,V$ plane for e-ph coupling, $g_{ep}=0.5$ with $\omega_{0}=1.0$. The phase diagrams show that three possible phases--CDW, SDW, and SS--are present, with the SS phase being restricted to small values of $U$ and $V$, as one would expect. For larger values of $U$ and $V$ and for these values of the phonon parameters, the model exhibits physics similar to that of the conventional extended Hubbard model. \cite{Tam_EH} We note that the delicate BOW phase mentioned in the introduction that occurs between the CDW and the SDW in the extended Hubbard model is not shown in our Fig. \ref{phase_diagram}.  From our previous studies \cite{Tam_EH} we know that this phase is not captured by the MFRG with $N_k=2$ (see Appendix), so we do not expect it to appear in our calculation. This choice was made deliberately to allow us to focus on the effects of e-ph interactions and phonon retardation, which are the phenomena that drive the superconducting fluctuations. As one would expect intuitively, the most interesting regime for SC is at weak e-e repulsion. In this region, we find the SS susceptibility unambiguously dominates. This is not what we see in the usual models with local e-ph interactions and this arises from the non-local nature of the e-ph interactions. 

\section{Analytic Insight into the gapless SC phase in the EHEH model}

In order to gain insight into the mechanism driving the dominant superconducting fluctuations in more detail, we next undertake an essentially analytic study that uses a standard renormalization group (RG) approach to the bosonized form of the EHEH model. For ease of presentation and comparison to the prior literature, we use the RG formulation of Yonemitsu and Imada \cite{Yonemitsu} to write the EHEH Hamiltonian entirely in terms of bosonic fields. Here we will keep only the lowest energy electrons--those around the Fermi points  at $k_{f}$ and $-k_{f}$--and the single phonon excitations that couple to the electrons (so the $2k_f$ and 0 momentum phonons). Because the difference between the EHEH model and previously studied models lies entirely in the electron-phonon interactions, we will focus first on their effects. Explicitly, this means that we ignore the back scattering and the Umklapp scattering from the direct e-e interactions and incorporate the remaining forward scattering into the Luttinger parameters. The purely electronic part is given by the standard free boson Hamiltonian, which we do not need here. With these simplifications, the e-ph interaction can be written as \cite{Voit}
\begin{eqnarray}
H^{ep} &=& H^{ep}_{1} + H^{ep}_{2}, \\
H^{ep}_{1} &=& \int dx \frac{1}{\pi \alpha} \{\gamma_{1}exp[\sqrt{2}i \Phi_{\rho}(x)]cos[\sqrt{2}\Phi_{\sigma}(x)]\nonumber\\
& & \phi_{2k_{f}}(x) + H.c.\} \nonumber\\ 
H^{ep}_{2}&=& \gamma_{2} \int dx [\rho_{+}(x) + \rho_{-}(x)] \phi_{0}(x), \nonumber 
\label{eph}
\end{eqnarray} 
Here $\gamma_{1} = (1+\exp(i2k_{f}))g_{ep}$, $\gamma_{2} = 2g_{ep}$; $\phi_{0}$, and $\phi_{2k_{f}}$ are the $k=0$ and the $k=2k_{f}$ components of the phonon fields respectively; $\Phi_{\rho}$, and $\Phi_{\sigma}$ are the bosonized fields for the charge part and the spin parts, respectively; $\rho_{+}$, and $\rho_{-}$ are the charge densities at the $k=k_{f}$, and $k=-k_{f}$ respectively; and $\alpha$ is the long distance cutoff. Note that in the HEH model, $\gamma_{1} = \gamma_{2} = g_{ep}$.
At this point, it is clear that at half-filling where $k_{f}=\pi/2$ the e-ph interaction contributes only to the linear forward scattering, $H_{2}$. The contribution from the non-linear term, $H_{1}$, vanishes exactly at half-filling. This is essentially the same result as shown in Eqn. (\ref{g}). Intuitively, this implies that the zero wave vector ordering benefits from the forward scattering but the $2k_{f}$ ordering, like that of the charge ordering, does not. Therefore, a charge gapless SC ordering is preferred to the charge gapped CDW ordering. The vanishing of $H_{1}$ at half-filling also means that the charge part and the spin parts are decoupled in the absence of direct e-e interactions. As a result of these peculiar features, this model can be diagonalized exactly, and it can be shown that the charge gap is zero. \cite{Loss} 

\begin{figure}[]
\centerline{
\includegraphics*[height=0.28\textheight,width=0.43\textwidth, viewport=0 0 360 300,clip]{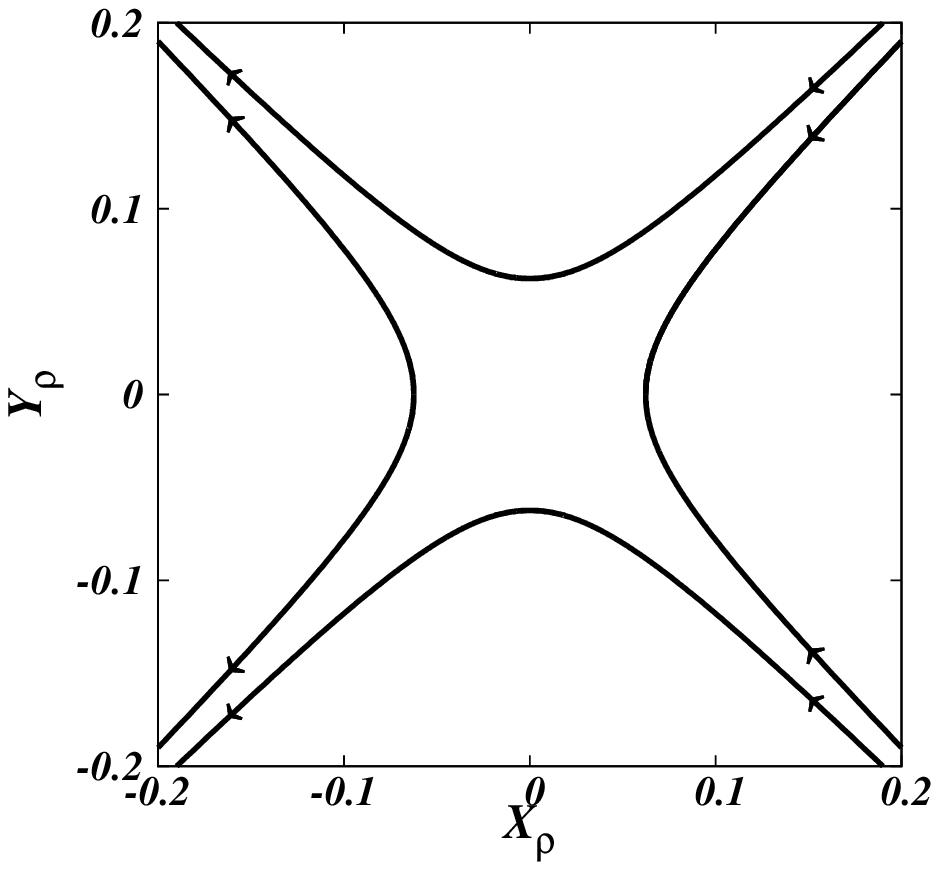}} 
\centerline{
\includegraphics*[height=0.28\textheight,width=0.43\textwidth, viewport=0 0 360 300,clip]{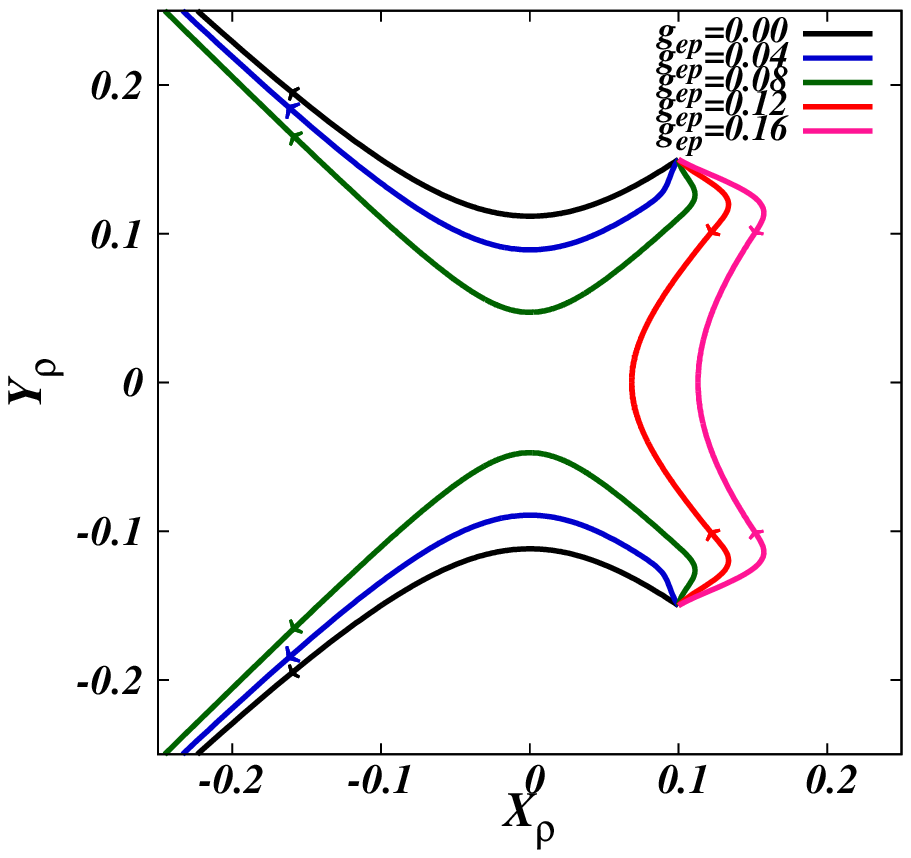}}
\caption{The flows of the charge coupling, $X_{\rho}$, and $Y_{\rho}$ without (the upper panel) and with electron-phonon couplings (the lower panel). See section IV for a detailed description.} 
\label{SG_flows}
\end{figure}

When we include direct e-e interactions, the bosonized Hamiltonian is no longer solvable. But the special feature of this model--namely, the spin-charge separation-- is preserved. Thus the spin part of the Hamiltonian will not be affected by the e-ph interaction which couples solely to the charge. 
In addition, as shown in Ref. [\onlinecite{Yonemitsu}], since the operator (proportional to $V$) for the spin-charge coupling in the extended Hubbard model has scaling dimension $4$, the charge part can be understood independently from the spin part for small coupling. Hence to illustrate the nature of gapless phase under the influence of both e-e and e-ph interactions, we employ the RG equations in the form used by Yonemitsu and Imada \cite{Yonemitsu} and focus solely on the flows of the {\it charge} couplings. As shown in Ref. [ \onlinecite{Yonemitsu}], these coupled flows are described by variables $X_{\rho} = 2(1-\kappa_{\rho}^{-1})$, where $\kappa_{\rho}$ is the usual Luttinger Liquid charge exponent and  $Y_{\rho}= g_3/(\pi v_f)$, where $g_3$ is the usual Umklapp coupling in g-ology. \cite{Voit} From the general form of Eqns. (3.11) and (3.12) in Ref. [\onlinecite{Yonemitsu}], we consider the special case of 1/2 filling and vanishing backward scattering from the phonons (see Eqn. \ref{eph}). In this case, the equations in Ref.[\onlinecite{Yonemitsu}] simplify considerably and become
\begin{eqnarray}
\frac{dX_{\rho}(l)}{dl} &=& -Y_{\rho}^{2}(l) + 2Y_{2}(l)D_{0}(l), \nonumber  \\
\frac{dY_{\rho}(l)}{dl} &=& -X_{\rho}(l)Y_{\rho}(l).
\end{eqnarray} 
Here the phonon propagator at momentum $0$ is $D_{0}=\frac{\omega_{0}}{E(l)}\exp(-\frac{\omega_{0}}{E(l)})$, with $E(l)=E_{f}\exp(-l)$. $E_{f}$ is the Fermi energy and $v_{f}$ is the Fermi velocity. The initial conditions are $X_{\rho}(0)=-(U+6V)/(\pi v_{f})$, $Y_{\rho}(0)=(U-2V)/(\pi v_{f})$, and $Y_{2}(0)=2g_{ep}^{2}/(\pi v_{f} \omega_{0}^{2})$. Notice that $Y_2$ does not flow with $l$. \cite{Yonemitsu} 
They are renormalization group equations for the sine-Gordon model, with the exception of an additional drift term from the forward scattering of the e-ph interaction scaled by the phonon propagator. The effect from the phonon propagator exists only roughly above the phonon energy. If one momentarily ignores the e-ph interaction term in the first equation, these equations can be integrated exactly to give ($X^2_{\rho} - Y^2_{\rho}) =C$, which shows that the RG trajectories are hyperbolas in the $X_{\rho}-Y_{\rho}$ plane. Further, there is a line of fixed points at $Y_{\rho}$=0 which are attractive (stable) for $X_{\rho} > 0$ and repulsive (unstable) for $X_{\rho} < 0$. Hence the RG flows are as shown in the upper panel of Fig. \ref{SG_flows}. In the quadrant bounded by $X_{\rho} = \pm Y_{\rho}$  and $X_{\rho}>0$, the flows go to a fixed point along the line $Y_{\rho}= 0$, which corresponds to the gapless phase in the sine-Gordon model and to the regime of dominant SC fluctuations in the EHEH model.  Outside this region, the equations always flow to strong coupling, reflecting instabilities toward the SDW ($Y_{\rho}  \rightarrow \infty,  X_{\rho} \rightarrow - \infty $) and CDW ($Y_{\rho}  \rightarrow -\infty, X_{\rho} \rightarrow -\infty $)  phases in the EHEH model. \cite{Nakamura2,Voit} In the lower panel of Fig. \ref{SG_flows}, the results of including the e-ph coupling are shown. Since the RG equation is now explicitly "time" (that is, scale) dependent from the phonon propagator, the RG flows not only depend on the initial values of the $X_{\rho}$ and $Y_{\rho}$, but also on the initial scale, $l=l_{0}$. Therefore, we cannot draw the RG trajectories as in the case without e-ph interaction, and we cannot solve the problem analytically. Instead, we use numerics and illustrate some typical flows near the boundary between the gapless and gapped phases as functions of $g_{\rm ep}$, with $\omega_0$ set to 1. The effect of the e-ph interaction is to "pull"  the flows towards a larger value of $X_{\rho}$. If $X_{\rho}$ is pulled far enough, it will eventually flow to a finite fixed point, while $Y_{\rho}$ will flow to zero (see fig. \ref{SG_flows}), and the system will be in the region of dominant SC fluctuations. Thus initial values that for small e-ph coupling flow to one of the density wave phases; for larger e-ph coupling flow instead to the superconducting phase. These approximated mostly analytic results thus corroborate our MFRG study which show that a charge gapless phase with dominant superconducting fluctuations can be obtained in the EHEH even with all repulsive e-e interactions where both $U$ and $V$ are positive.

\section{Conclusions}

We have shown that by incorporating non-local electron-phonon interactions we can find a 1D Hamiltonian--the EHEH model--that exhibits a gapless metallic phase with dominant superconducting fluctuations over a wide range of parameters. This contrasts with the results for models involving local electron-phonon interactions, where a gapless metallic phase, if it exists, does not appear to have dominant superconducting fluctuations and occurs over a limited parameter range. Although a true superconducting phase cannot exist in 1D, our results suggest that the inclusion of non-local e-ph interaction may be crucial to understanding the behavior of the quasi-two dimensional high-$T_c$ materials \cite{Zhou} and organic superconductors. \cite{Ishiguro1,Ishiguro2} Indeed, the recent experimental evidence that bosonic modes, very likely related to phonons, play a crucial role in some of the most interesting features of the high $T_c$ cuprate superconductors \cite{Zhou} provides further motivation for studying models with both e-e and non-local e-ph interactions. In this regard, we point out the recent study of the two-dimensional case for e-ph interactions beyond the local Holstein coupling where it has been shown that superconductivity does exist with some anisotropic non-local coupling even at half-filling; moreover, for this two-dimensional system, there is very likely to be a true long range SC order at zero temperature. \cite{Klironomos}

\section{Acknowledgements}
We thank R. Torsten Clay for many helpful discussions and are pleased to acknowledge the hospitality of the Aspen Center for Physics where some of this work was accomplished, and the Center for Computational Science at Boston University for partial support of the computational work, S.-W. Tsai gratefully acknowledges support from NSF under Grant No. DMR0847801 and from the UC-Lab FRP under award No. 09-LR-05-118602. during the completion of this work. 

\section{Appendix}

\subsection{The MFRG method and equations for the flows of susceptibilities}

\begin{figure}[bth]
\centerline{
\includegraphics*[height=0.28 \textheight,width=0.4\textwidth,viewport=0
0 360 280,clip]{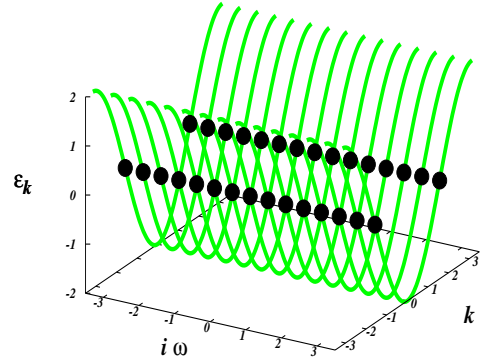}} \caption{Discretization of the momenta
in the Brillouin zone and frequencies in the frequencies axis. This figure shows the case $N_k=2, N_w =15$.}
\label{patches}
\end{figure}

The functional renormalization group for electron-electron coupled systems has been discussed extensively in the liteature. \cite{Shankar,Zanchi,Binz,Honerkamp,Salmhofer-book,Metzner-review,Honerkamp-review,Halboth,Tsai3}
The MFRG approach implementation \cite{Tsai1,Tsai2} at the one-loop level yields the following RG flow equations for the coupling {\it functions}, 
$g(k_1,k_2,k_3,k_4)$, with initial conditions as given by Eqn. \ref{g}  :
\begin{eqnarray}
&&\frac{dg(k_1,k_2,k_3)}{d \Lambda} =
\nonumber\\
&-&\!\!\!\!\int\!d p \frac{d}{d\Lambda}
[G_{\Lambda}(p)G_{\Lambda}(k)]
g(k_1,k_2,k) 
g(p,k,k_3)
\nonumber\\
&-&\!\!\!\!\int\!d p \frac{d}{d\Lambda}
 [G_{\Lambda}(p)G_{\Lambda}(q_1)]
g(p,k_2,q_1)
g(k_1,q_1,k_3) 
\nonumber\\
&-&\!\!\!\!\int\!d p \frac{d}{d\Lambda}
 [G_{\Lambda}(p) G_{\Lambda}(q_2)]
 [-\!2g(k_1,p,  
 q_2)g(q_2,k_2,k_3)
\nonumber\\
&+&\!\!g(p,k_1,q_2)
g(q_2,k_2,k_3)\!+
\!g(k_1,p,q_2) 
g(k_2,q_2,k_3)],
\label{eq:rg1}
\end{eqnarray}
where $k=k_1+k_2-p$,
$q_1=p+k_3-k_1$,
$q_2=p+k_3-k_2$, 
$\int d p=\int
d\vec{p}\sum_{\omega}1/(2\pi\beta)$, and $G_{\Lambda}$ is the self-energy corrected
propagator at energy cut-off $\Lambda$.
Since the interaction vertices are frequency dependent, there are also
self-energy corrections. At the one-loop level, the self-energy MFRG equation is:
\begin{eqnarray}
\frac{d \Sigma(k)}{d \Lambda} = 
&-&\!\!\!\!\int\!d p \frac{d}{d\Lambda} [G_{\Lambda}(p)]
[2g(p,k,k) -
g(k,p,k)]. 
\label{eq:rg2}
\end{eqnarray}
We have solved the coupled integral-differential equations,
(\ref{eq:rg1}) and (\ref{eq:rg2}), numerically with two
Fermi points ($N_k = 2$) and by dividing the frequency axis into fifteen segments ($N_\omega = 15$). 
Fig. \ref{patches} shows the discretization scheme for $N_k = 2$ and $N_\omega = 15$.  

We next calculate within the MFRG approach for the RG flow of susceptibilities in the  
static (zero frequency) and long-wavelength limit. 
The static susceptibilities for singlet superconductivity (SS) and triplet superconductivity (TS) are given by:
\begin{eqnarray}
\!\!\!\chi^{{\delta}}_{\Lambda}\!=\!\int \!D(1,2) \!f(p_1)f(p_2)
\langle c_{p_{1},\downarrow}c_{-p_{1},\uparrow}
c_{-p_{2},\uparrow}^{\dagger}c_{p_{2},\downarrow}^{\dagger}\!\rangle.
\end{eqnarray}
For $\delta={\rm SS}$, $f(p) =1$, whereas for $\delta={\rm TS}$, $f(p)=sin(p)$.

The static $2k_f$ charge density wave (CDW) and spin density wave (SDW) susceptibilities can be written as \cite{Tam_hh}:
\begin{eqnarray}
\chi^{\delta}_{\Lambda}\!\!=\!\!\!\int\! \!\!D(1,2)\!\!\!\sum_{\sigma_{1},\sigma_{2}}s_{\sigma_{1}}s_{\sigma_{2}}
\langle
c_{p_{1},\sigma_{1}}^{\dagger}c_{p_{1}\!+\!\pi,\sigma_{1}}
c_{p_{2}\!+\!\pi,\sigma_{2}}^{\dagger}c_{p_{2},\sigma_{2}} \rangle. 
\end{eqnarray}
For $\delta = {\rm SDW}$: $s_\uparrow =1, s_\downarrow =-1$, and for $\delta = {\rm CDW}$: $s_\uparrow =1, 
s_\downarrow = 1$. 

In the above equations, $p_i$ is the momentum at energy $\xi_{i}$,
$\int D(1,2)\equiv \int_{|\xi_{1}|>\Lambda}d\xi_{1}J(\xi_{1})
\int_{|\xi_{2}|>\Lambda}d\xi_{2}J(\xi_{2})$, 
and $J(\xi)$ is the Jacobian for the coordinate transformation
from $k$ to $\xi_k$. The dominant instability is determined by the most divergent susceptibility as the cut-off $\Lambda$ is 
lowered. 


The RG flows for the SS and TS susceptibilities are,
\begin{eqnarray} 
\label{SC_susceptibility RGE}
\!\!\!\!\!\!\!\!\frac{d \chi^{\delta}_{\Lambda}\!(0,0)}{d\Lambda}\!\!&=&
\!\!\int\!\!\!d p \frac{d}{d\Lambda}[
G_{\Lambda}(p)G_{\Lambda}(-p)]
(Z^{\delta}_{\Lambda}\!(p))^2, \\
\!\!\!\!\!\!\!\!\frac{d Z^{\delta}_{\Lambda}\!(p)}{d\Lambda} \!\!&=&\!
-\!\!\!\int\!\!\!d p^{\prime}\!\frac{d}{d\Lambda}[
G_{\Lambda}(p^{\prime})G_{\Lambda}(\!-p^{\prime})]
Z^{\delta}_{\Lambda}\!(p^{\prime})
g^{\delta}\!(p^{\prime},p), 
\end{eqnarray}
where $g^{\delta}(p^{\prime},p) =
g(p^{\prime}\!,- p^{\prime},
-p)$ for both $\delta=$SS and TS.

The RG flows for the SDW and CDW susceptibilities are,
\begin{eqnarray} \label{SDW_susceptibility RGE}
\!\!\!\!\!\frac{d \chi^{\delta}_{\Lambda}(\pi,0)}{d\Lambda}\!\!&=&
-\!\!\!\int\!\!\!d p \frac{d}{d\Lambda}[
G_{\Lambda}(p)G_{\Lambda}(p\! +
\!Q)](Z^{\delta}_{\Lambda}(p))^2, 
\\
\!\!\!\!\!\frac{d Z^{\delta}_{\Lambda}(p)}{d\Lambda} \!\!&=&
\!\!\!\int\!\!\!d p^{\prime}\frac{d}{d\Lambda}[
G_{\Lambda}(p^{\prime})
G_{\Lambda}(p^{\prime}\!\!+\!Q)] 
Z^{\delta}_{\Lambda}\!(p^{\prime})
g^{\delta}\!(p^{\prime}\!,p),
\end{eqnarray}
where $Q = (\vec{Q}=\pi, 0)$. For $\delta={\rm SDW}$:
$g^{\delta}(p^{\prime},p)= 
-g(p+Q,p^{\prime},p)$,
and for $\delta={\rm CDW}$ : 
$g^{\delta}(p^{\prime},p) =
2g(p^{\prime},p+Q,p) -
g(p +Q,p^{\prime},p)$.
The function $Z^{\delta}(p)$ is the effective vertex in the
definition of the susceptibility $\chi^{\delta}$. \cite{Zanchi}


\end{document}